\documentclass{nature}
\usepackage{graphicx} 
\usepackage{cite}
\usepackage{epsfig} 
\usepackage{amsmath}

\usepackage{xcolor}
\usepackage{soul}

\title{Observation of miniaturized bound states in the continuum with ultra-high quality factors}

\author{Zihao Chen$^{1}$, Xuefan Yin$^1$, Jicheng Jin$^2$, Zhao Zheng$^1$, Zixuan Zhang $^{1}$, Feifan Wang$^{1}$, Li He$^2$, Bo Zhen$^2$, \& Chao Peng$^{1,*}$}
\begin{document}
\maketitle

\begin{affiliations}
\item State Key Laboratory of Advanced Optical Communication Systems and Networks, Department of Electronics \& Frontiers Science Center for Nano-optoelectronics, Peking University, Beijing 100871, China
 \item Department of Physics and Astronomy, University of Pennsylvania, Philadelphia, PA 19104, USA

\end{affiliations}

\begin{abstract}
Light trapping is a constant pursuit in photonics because of its importance in science and technology. 
Many mechanisms have been explored,
including the use of mirrors made of materials or structures that forbid outgoing waves \cite{akahane_highq_2003,song_ultrahigh_2005,gondarenko_high_2009,lu_high_2014,biberman_ultralowloss_2012,vahala_optical_2003,srinivasan_cavity_2006,pernice_high_2012}, and bound states in the continuum that are mirror-less but based on topology \cite{hsu_observation_2013,lee_observation_2012,bulgakov_topological_2017,hsu_bound_2016,jin_topologically_2019,gao_diracvortex_2020}.
Here we report a compound method, combing mirrors and bound states in the continuum in an optimized way, to achieve a class of on-chip optical cavities that have high quality factors and small modal volumes. Specifically, light is trapped in the transverse direction by the photonic band gap of the lateral hetero-structure and confined in the vertical direction by the constellation of multiple bound states in the continuum. As a result, unlike most bound states in the continuum found in photonic crystal slabs that are de-localized Bloch modes, we achieve light-trapping in all three dimensions and experimentally demonstrate quality factors as high as $Q = 1.09 \times 10^6$ and modal volumes as low as $V = 3.56~ \mu m^3$ in the telecommunication regime. We further prove the robustness of our method through the statistical study of multiple fabricated devices. Our work provides a new method of light trapping, which can find potential applications in photonic integration\cite{fang_ultracompact_2011,tanabe_trapping_2007,munzberg_superconducting_2018,ellis_ultralowthreshold_2011,sun_singlechip_2015,hirose_wattclass_2014,yu_demonstration_2017,yao_onchip_2010}, nonlinear optics \cite{soljacic_enhancement_2004,carletti_giant_2018,bernhardt_quasibic_2020,minkov_doubly_2019,koshelev_subwavelength_2020} and quantum computing\cite{hacker_photon_2016,liu_singlephoton_2019,choi_selfsimilar_2017}. 
\end{abstract}

While most light-trapping methods rely on the use of mirrors to forbid radiation, it is recently realized that optical bound states in the continuum (BICs) provide an alternative approach. 
BICs are localized wave functions with energies embedded in the radiation continuum, but, counter-intuitively, do not couple to the radiation field. 
So far, BICs have been demonstrated in multiple wave systems, including photonic, phononic, acoustic, and water waves, and found important applications in surface acoustic wave filters and lasers\cite{meier_laser_1998,imada_coherent_1999,hirose_wattclass_2014,hsu_observation_2013,lee_observation_2012,tong_observation_2020,cobelli_experimental_2009,cumpsty_excitation_1971,koshelev_subwavelength_2020}.
In many cases, BICs can be understood as topological defects\cite{kitamura_focusing_2012,iwahashi_higherorder_2011,zhang_observation_2018,doeleman_experimental_2018,lu_topological_2014}: for example, they are fundamentally vortices in the far field polarization in photonic crystal (PhC) slabs, each carrying an integer topological charge\cite{zhen_topological_2014}. 
Manipulations of these topological charges have led to interesting consequences, including resonances that become more robust to scattering losses\cite{jin_topologically_2019} and unidirectional guided resonances that only radiate towards a single side without the use of mirrors on the other\cite{yin_observation_2020}. 

So far, most BICs studied in PhC slabs are only localized in the vertical (thickness) direction, but remain de-localized in the transverse direction across the slab, rendering them less ideal in enhancing light-matter interaction with localized emitters or quantum applications. While it is known that perfect BICs localized in all three dimensions cannot exist \cite{silveirinha_trapping_2014}, it is of great interest to explore the limit of BIC miniaturization.  
A simple truncation of the PhC slab can reduce modal volume $V$, but also drastically degrades the quality factor $Q$, as it introduces leakage in both lateral and vertical directions. 
A common relationship between $Q$ and $V$ for BICs with out-going boundary conditions in plane has been derived \cite{chua_lowthreshold_2011}, showing good agreements with experiments\cite{liang_threedimensional_2012}. 

Here we theoretically propose and experimentally demonstrate a method to achieve miniaturized BICs (mini-BICs) in PhC slabs, through the proper arrangement of multiple topological charges in the momentum space.  
Specifically, we start by enclosing the mini-BIC with a photonic band-gap mirror, using a lateral hetero-structure, to forbid transverse leakage.  
Similar to electronic quantum dots, the continuous photonic bands of an infinite PhC turn into discrete energy levels, due to the momentum quantization according to the size of the mini-BIC.
For the same reason, the out-of-plane leakage of the mini-BIC is also dominated by a few directions that satisfy the momentum-quantization condition. 
As the PhC unit cell design is varied,  multiple BICs\cite{ni_tunable_2016,yang_analytical_2014,hsu_observation_2013,lee_observation_2012}--- each carrying a topological charge and together composing a topological constellation --- evolve in the momentum space, and eventually match with the major leakage channels. 
At this point, the out-of-plane radiation of the mini-BIC is topologically eliminated, giving rise to an ultra-long lifetime and a small modal volume. 

{\bf Design and topological interpretation |} As a specific example, we consider a PhC slab (Fig. 1a), where circular air holes  (radius $r=175$ nm)  are defined in a silicon layer of $h=600$ nm thick. 
The mini-BIC design consists of a square cavity region $A$ surrounded by a boundary region $B$ with a gap size of $g=541$ nm in between. 
The cavity $A$ has a length of $L$ in each side. 
Region $A$ and $B$ have different periodicities, $a=529$ nm and $b=552$ nm, to form a heterostructure in-plane. 
We focus on the lowest-frequency TE band in region $A$ above the light line (black line in Fig.~1b), whose energy is embedded in the bandgap of region $B$. 
By tuning parameters $b$ and $g$, the interface between region $A$ and $B$ is almost perfectly reflective, minimizing the lateral leakage of the mini-BIC. 
In a finite sized structure, the continuous band of a PhC (left panel of Fig.~1b) splits into a series of discrete modes (right-panel) as the continuous momentum space is quantized into isolated points with a spacing of $\delta k = \pi/L$ in between (Fig. 1c). 
This is analogous to to what happens in an electronic quantum dot.
Each mode can thus be labelled by a pair of integers $(p,q)$, indicating that its momentum is mostly localized near $(p \pi/L, q \pi/L)$ in the first quadrant. 
Four modes, M$_{11}$ through M$_{22}$, are shown as examples, where M$_{12}$ and M$_{21}$ are degenerate in frequency due to the 90-degree rotation symmetry of the structure ($C_4$). 
These modes exhibit distinctly different near- and far-field patterns (Fig. 1d), which are determined by their quantized momenta accordingly. 
The modal volumes of M$_{11}$ through M$_{22}$ are calculated as $3.56~\mu$m$^3$, $3.37~\mu$m$^3$ and $3.26~\mu$m$^3$, respectively.
More details on the theory and design are presented in Section \uppercase\expandafter{\romannumeral1} through \uppercase\expandafter{\romannumeral5} of the Supplementary Information. 

Next, we show that the radiation loss of each mode can be strongly suppressed through the topological constellation of BICs.  
Fundamentally, BICs are topological defects in the far-field polarization, which carry integer topological charges:
\begin{equation}
    q = \oint_C{d\mathbf{k}\cdot\nabla_\mathbf{k}\phi(\mathbf{k})}.
\end{equation}
Here $\phi(\mathbf{k})$ is the angle between the polarization major axis of radiation from the mode at $\mathbf{k}$ and the $x-$axis. 
$C$ is a simple closed path that goes around the BIC in the counter-clockwise (CCW) direction. 
As shown in Fig.~2a, for an infinite PhC with $a=526.8$ nm (case $W$), there are 9 BICs: one is at the center of the Brilluion zone (BZ), and the other 8 form an octagonal-shaped topological constellation, which is denoted by their distance to the BZ center ($k_{\text{BIC}}$). 
The position of the topological constellation can be controlled by varying the periodicity $a$: for example, as $a$ increases from $526.8$ nm (case $W$ in Fig.~2a) to $534$ nm (case $Z$), the topological constellation shrinks and merges together before it turns into a single topological charge. 
The evolution of the $Q$s of infinite PhCs is shown in the lower panel of Fig.~2a.

Whenever the topological constellation matches with the main momenta of a mode $M_{pq}$, i.e. $k_{\text{BIC}} L/\pi= \sqrt{p^2+q^2}$, its radiation loss is strongly suppressed, as its major underlying Bloch mode components are now BICs with infinitely high $Q$s. This is confirmed by our simulation results in Fig. 2b (see Methods for more details): the $Q$ of M$_{11}$ (red line) is maximized in case $X$ when the matching condition is met. The maximum $Q$ exceeds $8\times 10^6$.
Similarly, the $Q$s of M$_{12}$ (blue) and M$_{22}$ (black) are also maximized when the matching conditions are satisfied, labelled by blue and black dashed lines, respectively.
Here we note that all localized modes penetrate, partly, into the boundary region, so the effective cavity length $L_{\rm eff}$ is calculated as $22.3a$, which is slightly larger than the physical length of the cavity $L = 17a$.
More details on the simulation are presented in the Section \uppercase\expandafter{\romannumeral6} of the Supplementary Information.  

{\bf Sample fabrication and experimental setup | } To verify our theoretical findings, we fabricate PhC samples using e-beam lithography and induced coupled plasma etching processes on a 600 nm thick silicon-on-insulator wafer (see Methods for more details on the fabrication). The scanning electron microscope images of the samples are shown in Fig.~3.  The underlying SiO$_2$ layer is removed before measurements to restore the up-down mirror symmetry, required by the off-normal BICs. 
The footprint of the each sample is about $19.8~\mu$m $\times$ $19.8~\mu$m, including a cavity region that is $11.9~\mu$m $\times$ $11.9~\mu$m in size.
The periodicity of the cavity region $a$ is varied from $518$ to $534$ nm to satisfy the matching-condition and maximize $Q$s
 of M$_{11}$, M$_{12}$ and M$_{22}$. 
The periodicity of the boundary region and the gap distance are fixed at $b=552$ nm and $g=541$ nm. 

The experimental setup is schematically shown in Fig.~3d, which is similar to our previously reported results \cite{jin_topologically_2019,yin_observation_2020}. 
A tunable laser in the telecommunication band is first sent through a polarizer in the $y$-direction (POL Y) before it is focused by a lens (L1) onto the rear focal plane (RFP) of an infinity-corrected objective lens. 
The incident angle and beam diameter of the laser are fine tuned by L1 to maximize the coupling efficiency. 
The reflected beam is collected by the same objective, and further expanded by 2.67 times through a $4f$ system to best fit the camera.  
A X-polarizer (Pol X) is used to block direct reflection from the sample, while allowing the resonance's radiation to pass. 
See Methods for more details on the experimental setup.

{\bf Experimental results |} 
Whenever the excitation laser wavelength becomes on-resonance with a mode, the scattered light from the sample, captured on the camera, is maximized, which allows us to measure the resonance frequencies and $Q$s of different modes. 
Furthermore, under on-resonance condition, the far-field radiation pattern of each mode can also be recorded by the camera with polarizers (Fig.~4a). 
In particular, the far-field radiation of mode M$_{11}$, M$_{12}$ and M$_{22}$ is measured to be donut-, dipole- and quadrupole-shaped, respectively, showing good agreements with numerical simulation.   
Furthermore, by placing a pin hole (not show in Fig.~3d) at the image plane of the RFP of the objective to reject stray light, scattered light intensity is recorded using a photo-diode as the wavelength of the tunable laser is scanned. 
Distinct and sharp resonance peaks are found (mid-panel in Fig.~4b), corresponding to the 4 modes, M$_{11}$ through M$_{22}$. 
We note that the fabrication imperfection slightly breaks the $C_4$ symmetry and causes a minimal energy difference ($\approx$0.03\%) between M$_{12}$ and M$_{21}$ 

Higher resolution measurements near each mode yield results shown in the left and right panels of Fig.~4b. The design is optimized for mode M$_{11}$ ($a=529$ nm). 
The $Q$ of each mode is extracted by numerically fitting the scattering spectrum to a Lorentzian function.  
As shown, the measured $Q$s of modes M$_{12}$ and M$_{22}$ are $3.36 \times 10^5$ and $1.61 \times 10^5$, respectively. Meanwhile, the highest $Q$ for mode M$_{11}$ reaches $1.09 \times 10^6$, corresponding to a full width half maximum of $1.44$ pm.
To the best of knowledge, this is a record-high quality factor in small modal volume of BICs, which is about 60-fold enhancement of $Q$ and 4-fold shrinking of $V$ comparing with previously reported results\cite{liu_high_2019}.

Furthermore, to demonstrate the suppression of radiation loss originates from topological constellation, we vary periodicity $a$ between $518$ and $534$ nm and track how $Q$ changes. 
The measured wavelength of all modes agree well with simulation results (Fig.~5a). 
We see that, indeed, their measured $Q$s are always maximized when the topological-constellation-matching condition is met, which happen when $a =529.1$, $525.9$, and $522.8$ nm for M$_{11}$, M$_{12}$ and M$_{22}$, respectively (Fig. 5b). 
This finding shows good agreement with our simulation results in Fig.~2b. 
Finally, we prove the robustness of our method by measuring 87 different samples fabricated under the same design and through the same process.  
The histogram of their measured $Q$s of mode M$_{11}$ is shown in Fig. 5c, featuring an averaged $Q$ of $6.65 \times 10^5$ with a modest standard deviation of $1.22 \times 10^5$. See Supplementary Information Section \uppercase\expandafter{\romannumeral7} to \uppercase\expandafter{\romannumeral9} for more details.

To summarize, we present a type of  ultra-high-$Q$ and ultra-compact  mini-BICs by combining in-plane mirrors and out-of-plane BICs in an optimized way. 
We experimentally demonstrate a record-high quality factor for BICs of $Q=1.09 \times 10^6$ and a small modal volume of $3.56 ~\mu$m$^3$. 
Our finding can potentially lead to on-chip lasers with ultra-low thresholds\cite{ellis_ultralowthreshold_2011,wu_monolayer_2015,yao_onchip_2010}, chemical or biological sensors\cite{zhen_enabling_2013,romano_labelfree_2018,yanik_seeing_2011,chen_exceptional_2017}, nonlinear nanophotonic devices\cite{soljacic_enhancement_2004,carletti_giant_2018,bernhardt_quasibic_2020}, and elements for quantum computing\cite{hacker_photon_2016,liu_singlephoton_2019,choi_selfsimilar_2017}. 
Furthermore, our method of achieving ultra-high-$Q$ and ultra-low-$V$ are proven to be robust, owing to their topological nature, which paves the way to further improving the performance of optoelectronic devices.

\begin{methods}
\setcounter{equation}{0}
\renewcommand{\theequation}{M\arabic{equation}}

\subsection{Numerical simulation} 
All simulations are performed using the COMSOL Multiphysics in the frequency domain. Three-dimensional models are created with photonic crystal slabs placed between two perfect-matching layers. In other words, we have periodic boundary condition in-plane and out-going boundary condition in the vertical direction.  
The spatial meshing resolution is adjusted to adequately capture resonances with $Q$s of up to $10^9$. 
The eigenvalue solver is used to compute the frequencies and the quality factors of the resonances. 
The far-field emission patterns are computed by first retrieving the complex electric fields $\textbf{E}_{0,j}$ ($j=x,y$) just above the PhC surface and then calculating the emission fields as: 
\begin{equation}
    F_j(\theta, \phi) \propto (\cos \theta + \cos \phi -1) \iint_{x,y}E_{0,j}(x,y) e^{-ik_0(\tan \theta x+ \tan \phi y )}dx dy.
\end{equation}

\subsection{Sample fabrication.}
We fabricate the sample on a silicon-on-insulator (SOI) wafer with e-beam lithography (EBL) followed by induced coupled plasma (ICP) etching. 
For EBL, we first spin-coat the cleaved SOI chips with a 500nm-thick layer of ZEP520A photo-resist before it is exposed with EBL (JBX-9500FS) at beam current of $400$ pA and field size of 500 $\mu$m. 
Then we etch the sample with ICP (Oxford Plasmapro Estrelas 100) using a mixture of SF$_6$ and CHF$_3$. 
After etching, we remove the resist with N-Methyl-2-pyrrolidone (NMP) and the buried oxide layer using $49\%$ HF.

\subsection{Measurement setup.} We use a tunable laser (Santec TSL-550, C+L band) to generate  incident light. The light is first sent through a polarizer (Y-Pol) and is focused by a lens (L1) onto the rear focal plane an objective (Mitutoyo Apo NIR, 50X). The reflected and scattered light was collected by the same objective and a 4$f$ system is used to adjust the magnification ratio to 2.67X to best fit the observation. After passing through an orthogonal polarizer (X-Pol), only the scattered light is collected using a photo-diode (PDA10DT-EC). The resonance peaks are recorded by a high-speed data acquisition card (NI PCIe-6361)  connected to the photo-diode during wavelength scanning, and then fitted to Lorentzian function. A flip mirror is used to switch between the camera (PI NIRvana) and photodetector, in order to record the $x$-polarized far-field pattern of each mode. The $y$-polarized far-field pattern can similarly be obtained by switching the two polarizers (Y-Pol/X-Pol) to their orthogonal polarized directions (X-Pol/Y-Pol), respectively. We further obtained the overall far-field pattern by combining the $x$- and $y$-polarized patterns together.
Besides characterizing far-field patterns, the setup could also switch to near-field observation if another lens is inserted between L2 and L3.
\end{methods}

\bibliography{MiniBIC}{}
\bibliographystyle{naturemag}

\begin{addendum}
 \item[Correspondence] Correspondence and requests for materials
should be addressed to Chao Peng.~(email: pengchao@pku.edu.cn).
\end{addendum}

\clearpage
\begin{figure}
\centering
\includegraphics[width=10cm]{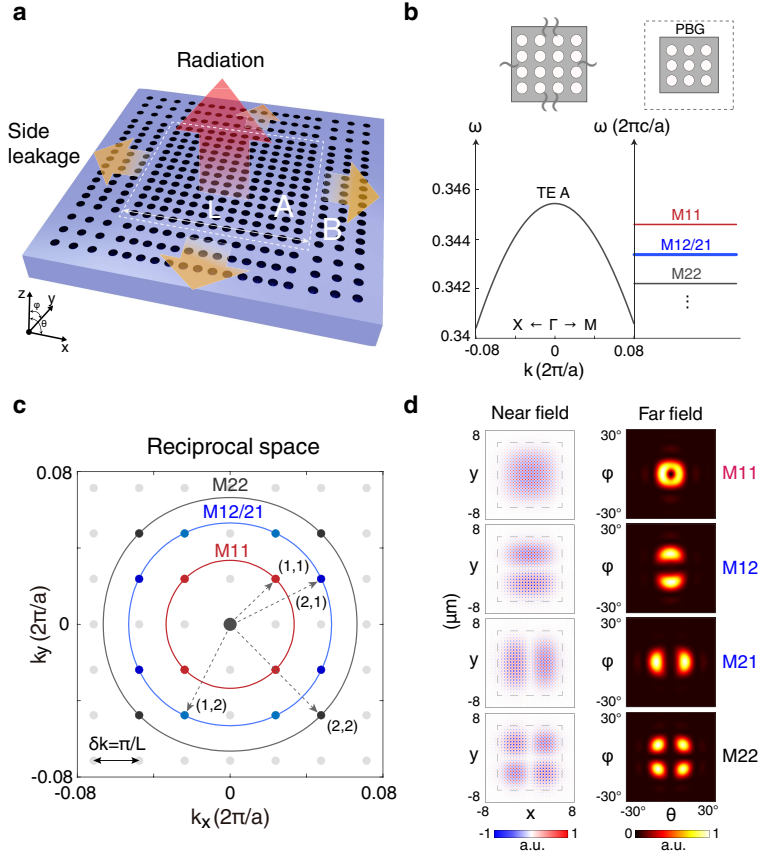}
\caption{{ $\mid$ \bf Mini-BIC modes.
} 
{\bf a,} Schematic of a mini-BIC (region A)
surrounded a photonic bandgap (PBG, region B). 
{\bf b,} 
A continuous band (TE-A) of an infinitely large PhC with periodic boundary condition (left) turns into a set of discrete modes under the PBG boundary condition (right).  
{\bf c,} 
The momentum distribution of each mode is highly localized to points that form a square lattice in the momentum space with a spacing of $\pi/L$.
Modes are labeled as M$_{pq}$, according to their momentum peak positions in the first quadrant at $(p\pi/L, q\pi/L)$. 
{\bf d,} 
The near-field mode profiles of four modes M$_{11}$ through M$_{21}$(left) and their far-field emission patterns (right). 
}
\end{figure}

\clearpage
\begin{figure}
\centering
\includegraphics[width=10cm]{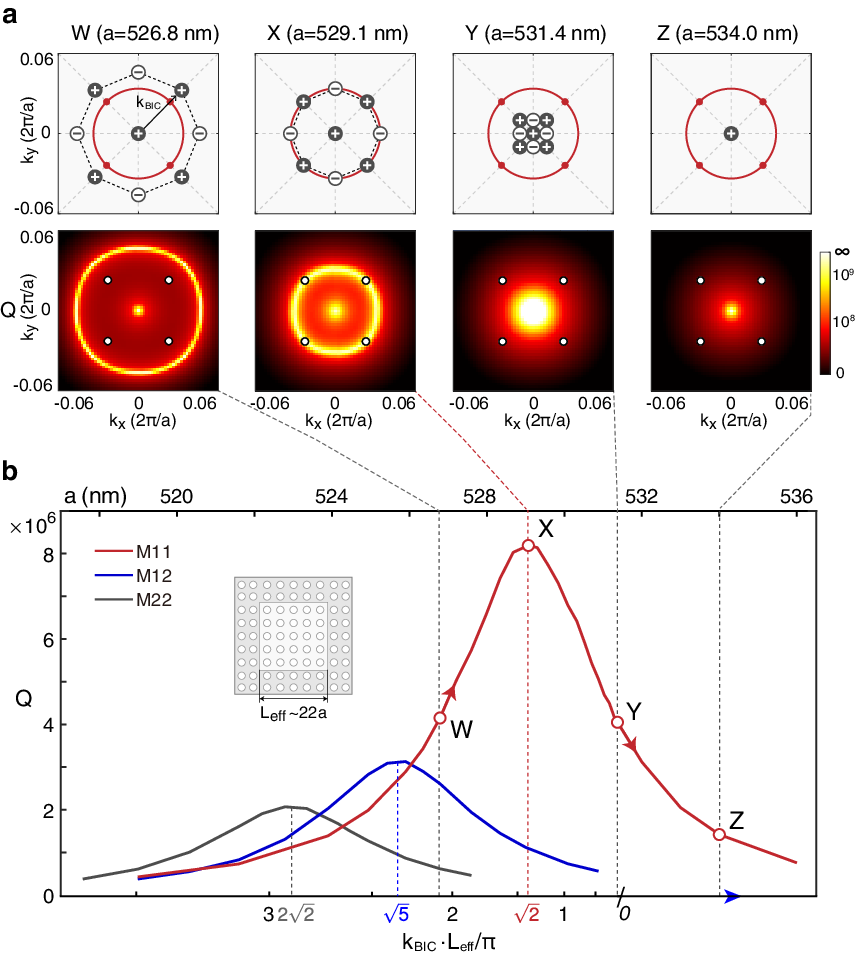}
\caption{{ $\mid$ \bf 
Maximizing the $Q$s of mini-BICs by properly arranging topological charges in the momentum space.
} 
{\bf a,} 
Multiple BICs appear on bulk band TE-A in momentum space, in which 8 off-$\Gamma$ ones with $q=\pm 1$ topological charges compose an octagonal-shaped topological constellation, denoted by the radius $k_{\text{BIC}}$.
When unit cell periodicity $a$ varies from $526.8$ nm (W) to $534.0$ nm (Z), the topological constellation shrinks, merges, and annihilates to a single topological charge (upper panel). 
The quality factor $Q$ for each unit cell design is shown in the lower panel.
{\bf b,} 
The quality factor $Q$ of  modes, $M_{11}$ through $M_{22}$, as functions of periodicity $a$ (upper axis) and topological constellation $k_{\text{BIC}}$ (lower axis). 
$Q$ for M$_{11}$ (red line) maximizes when its quantized momentum $\sqrt{2} \pi/L$ matches the topological constellation $k_{\text{BIC}}$, corresponding to case X ($a = 529.1$nm) in \textbf{a}. 
Similar maxima are also observed for M$_{12}$ (blue) and M$_{22}$ (black) under other designs, when $k_{\text{BIC}}$ matches $\sqrt{5} \pi/L$ and $2\sqrt{2} \pi/L$, respectively. 
}
\end{figure}

\clearpage
\begin{figure}
\centering
\includegraphics[width=10cm]{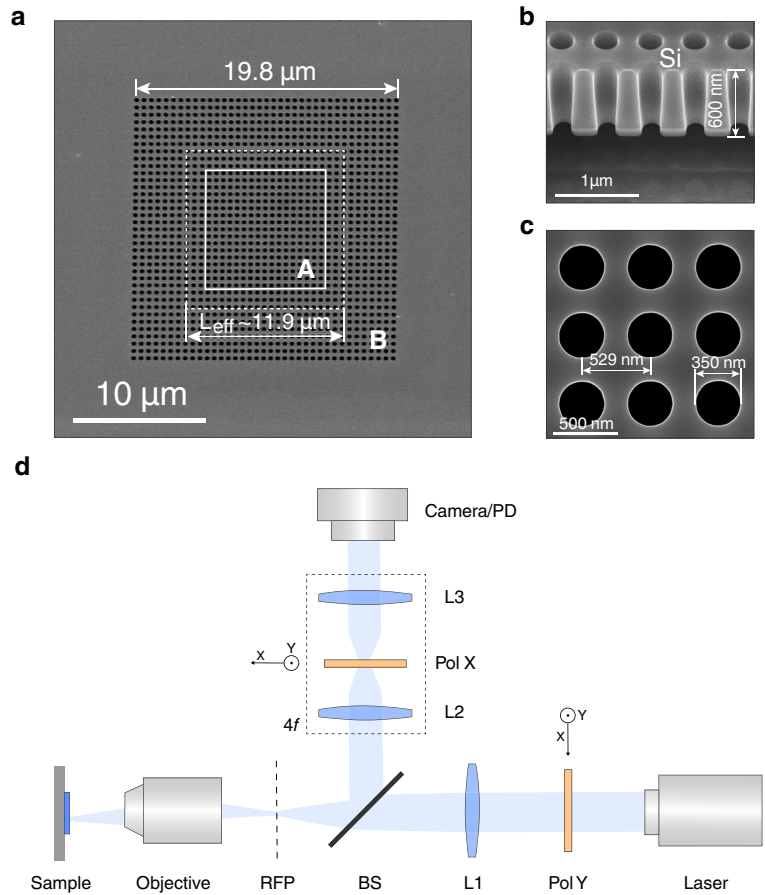}
\caption{{ $\mid$ \bf Fabricated sample and experimental setup.} {\bf a,b,c} Scanning electron microscope images of the fabricated samples from top and side views.  
The underlying SiO$_2$layer is removed before measurements. 
The chosen structural parameters correspond to case X in Fig. 2a to maximize $Q$ for mode $M_{11}$. 
{\bf d,} Schematic of the experimental setup. L, lens;  RFP, real focal plane; PD, photodiode; POL, polarizer; BS, beam-splitter; Lens L2 and L3 are confocal.}
\end{figure}

\clearpage
\begin{figure}
\centering
\includegraphics[width=14cm]{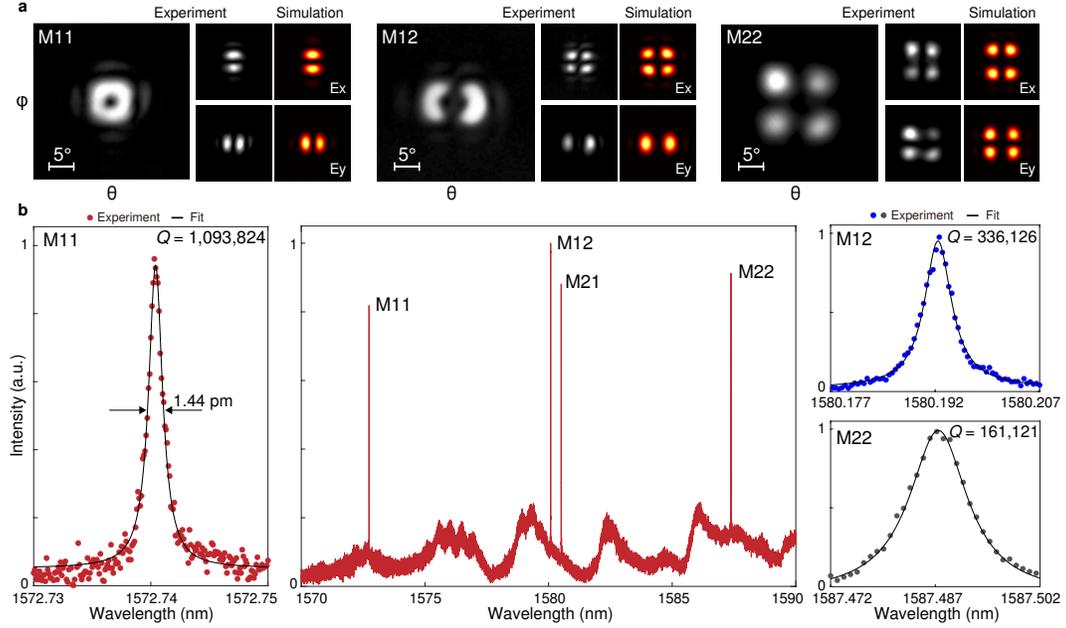}
\caption{{ $\mid$ \bf Observation of 
mini-BIC modes.} {\bf a,} The far-field emission patterns ($x$-, $y$-polarized and overall) of modes M$_{11}$ through M$_{22}$, measured with a camera (gray color map), show good agreements with simulation results (hot color map). 
{\bf b,} 
Middle panel: measured scattered light intensity as the laser wavelength scans from $1570$ nm to $1590$ nm. 
Four clear peaks are observed and identified as M$_{11}$ through M$_{22}$.
The $Q$ of $M_{11}$ reaches $1.09\times 10^6$ (left panel).  
In the same sample, the $Q$s of M$_{12}$ and M$_{22}$ are measured as $3.36 \times 10^5$ and $1.61 \times 10^5$, respectively (right panel).
}
\end{figure}

\clearpage
\begin{figure}
\centering
\includegraphics[width=14cm]{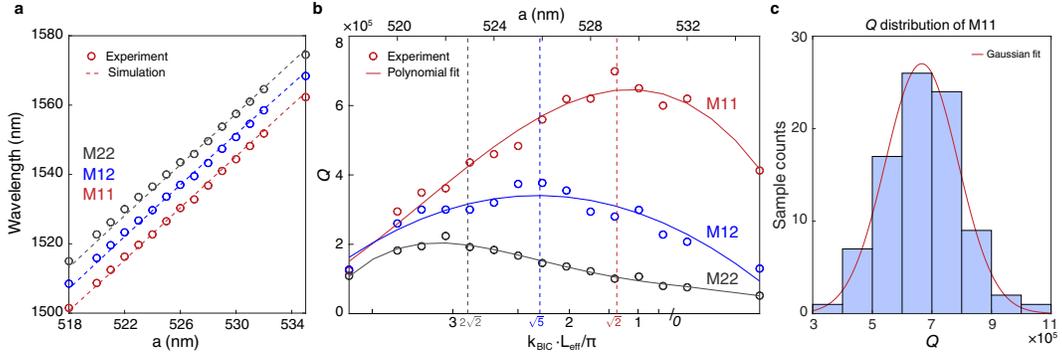}
\caption{{ $\mid$ \bf Demonstration of mini-BIC robustness against fabrication errors.
} 
{\bf a,} Measured resonance wavelengths (circles) in samples with different periodicities $a$ show good agreements with simulation results (dashed lines).
{\bf b,} 
Measured $Q$s (circles) in samples with different periodicities $a$ (upper axis) and, therefore, different BIC constellation ($k_{\rm BIC}$, lower axis). 
Polynomial fittings are shown in solid lines. 
Each curve is maximized when the matching condition is satisfied, indicated as dashed vertical lines. 
{\bf c,} Histogram statistics of measured $Q$s of M$_{11}$ in 87 samples, showing the robustness of our mini-BICs.
}
\end{figure}

\end{document}